# The Finite Temperature Phase Transition in the Lattice SU(2)-Higgs Model

K. Farakos[a], K. Kajantie[b], K. Rummukainen[c] and M. Shaposhnikov[d]

[a]National Technical University of Athens, Physics Department, Zografou, Gr-15780, Athens, Greece

[b]Department of Theoretical Physics, P.O.Box 9, 00014 University of Helsinki, Finland

[c]Indiana University, Department of Physics, Swain Hall-West 117, Bloomington, IN 47405, USA

[d]Theory Division, CERN, CH-1211 Geneva 23, Switzerland

We study the finite temperature transition of SU(2)-Higgs model with lattice Monte Carlo techniques. We use *dimensional reduction* to transform the original 4-dimensional SU(2)-gauge + fundamental Higgs theory to an effective 3-dimensional SU(2) + adjoint Higgs + fundamental Higgs model. The simulations were performed with Higgs masses of 35 and 80 GeV; in both cases we observe a stronger first order transition than the perturbation theory predicts, indicating that the dynamics of the transition strongly depend on non-perturbative effects.

## 1. Introduction

Recently proposed scenarios of the electroweak baryogenesis [1] are strongly dependent on the first order nature of the electroweak phase transition. Perturbative calculations indicate that for moderate Higgs boson masses the transition is of first order, albeit rather weakly; however, recent two-loop effective potential calculations give somewhat stronger first order transition [2]. In order to study non-perturbative effects lattice simulations are needed. Because the transition is expected to be rather weak, the Higgs mass at the transition temperature $T_c$ is much smaller than $T_c$, requiring very large lattices in order to avoid too severe finite size effects.

In this work we use *dimensional reduction* to reduce the 4-dimensional theory into a 3-dimensional *effective action*, with perturbatively calculable coefficients. This 3-d action is then studied with non-perturbative lattice simulations. This approach has some major advantages over the conventional 4-dimensional simulation: first, the 3-dimensional lattice obviously has considerably less variables than the 4-dimensional one with the same spatial volume, making it possible to simulate larger volumes. Also, all the timelike non-constant modes are treated analytically; these modes account for Debye screening, which is perturbatively rather well described at high temperatures. By excluding these modes from numerical simulation one can expect some reduction in the ultraviolet noise. On the other hand, the 3-dimensional non-perturbative physics remains intact.

We report numerical studies performed with two zero-temperature Higgs masses: $m_H = 35$ and 80 GeV, with three different lattice spacings on lattices up to $32^3$. Although unrealistically low, $m_H = 35$GeV produces a stronger first order transition than 80GeV, making it easier to identify the transition. These calculations are described in detail elsewhere [3,4]. There also exists a full 4-dimensional lattice calculation by Bunk et al. [5]. The dimensional reduction of the high-temperature SU(2) and SU(3) lattice gauge theories has been studied by Lacock et al. [6].

## 2. Effective Action

We start from the 4d SU(2) gauge + fermion + fundamental Higgs model and transform it into a 3d SU(2) gauge + fundamental Higgs + adjoint Higgs effective theory, the adjoint Higgs arising from the timelike component of the gauge field $A_0$. Suppressing all indices, the effective action is, up to 1-loop corrections,

$$S_{\rm eff} = \frac{1}{T}\int {\rm d}^3x\Big\{\frac{1}{4}F^2 + \frac{1}{2}(DA_0)^2 + (D\phi)^\dagger(D\phi)$$
$$+\frac{1}{2}\Big[\frac{1}{6}(5+N_F)g^2T^2 - 5g^2T\Sigma_c\Big]A_0^2$$
$$+\frac{g^4}{192\pi^2}(17-2N_F)A_0^4 + \Big[-\frac{1}{2}m_H^2 + (\frac{3}{16}g^2 + \frac{1}{2}\lambda$$
$$+\frac{g^2 {m_{\rm top}}^2}{8m_W^2})T^2 - (\frac{9}{4}g^2 + 6\lambda)T\Sigma_c\Big]\phi^\dagger\phi + \lambda(\phi^\dagger\phi)^2$$
$$+\frac{1}{4}g^2A_0^2\phi^\dagger\phi\Big\}. \qquad (1)$$

Note that the fermionic fields vanish completely; 4d fermions act only by renormalizing the 3d couplings. Here $\Sigma_c$ is the cutoff-dependent linearly divergent integral

$$\Sigma_c \equiv \int \frac{{\rm d}^3p}{(2\pi)^3}\frac{1}{\mathbf{p}^2}. \qquad (2)$$

We fix the parameters of the effective action by taking $g = 2m_W(\sqrt{2}G_F)^{-1/2} = 2/3$ and $m_W = 80.6$GeV; then $\lambda/g^2 = m_H^2/(8m_W^2)$. In the following, we also take $N_F = 0$.

We use the notations $A_0 = iA_0^a\sigma^a$, $\Phi = (\phi_0 + i\sigma_i\phi_i)/\sqrt{2}$, and reparametrize

$$igaA_0 \to A_0, \qquad \Phi \to \sqrt{\frac{T\beta_H}{2a}}\Phi\,. \qquad (3)$$

Using the shorthand notation $A_0^2 = \frac{1}{2}{\rm Tr}\,A_0^2$, $\Phi^2 = \frac{1}{2}{\rm Tr}\,\Phi^\dagger\Phi$, the action becomes

$$S = \sum_x\Big\{\beta_G\sum_{i<j}(1-\frac{1}{2}{\rm Tr}\,P_{ij}) +$$
$$+\beta_G\sum_i[\frac{1}{2}{\rm Tr}\,A_0(x)U_i^{-1}(x)A_0(x+i)U_i(x) - A_0^2]$$
$$+[10\Sigma(N^3) - \frac{4}{3}\frac{5}{g^2\beta_G}]A_0^2 + \frac{17}{16}\frac{g^2\beta_G}{3\pi^2}(A_0^2)^2$$
$$-\beta_H\sum_i[\frac{1}{2}{\rm Tr}\,\Phi^\dagger(x)U_i(x)\Phi(x+i)] +$$
$$+(1-2\beta_R)\Phi^2 + \beta_R(\Phi^2)^2 - \frac{1}{2}\beta_HA_0^2\Phi^2\Big\}, \qquad (4)$$

where $N$ is the linear length of the lattice, and

$$\Sigma(N^3) = \frac{1}{4N^3}\sum_{n_{1,2,3}=0}^{N-1}\frac{1}{\sum_{j=1}^3\sin^2(\pi n_j/N)}, \qquad (5)$$

where the term with all $n_i = 0$ is omitted. The relations between $g$, $T$ and $\lambda$ and the lattice couplings $\beta_G$, $\beta_H$, $\beta_R$ and the lattice spacing $a$ are given by

$$\begin{aligned}\beta_G &= \frac{4}{g^2}\frac{1}{Ta}\\ \beta_R &= \frac{1}{4}\lambda Ta\beta_H^2 = \frac{m_H^2}{8m_W^2}\frac{\beta_H^2}{\beta_G} \qquad (6)\\ \mu^2(T) &= \frac{2(1-2\beta_R-3\beta_H)}{\beta_Ha^2},\end{aligned}$$

where $\mu^2(T)$ is the coefficient of the $\phi^\dagger\phi$-term in the action eq. (1), with the replacement $\Sigma_c \to \Sigma(N^3)/a$. Since there are only 3 continuum parameters, the eqs. (6) define 1-dimensional curves of constant physics in the 4-dimensional lattice parameter space.

## 3. Simulations and Results

For concreteness, we fixed $g = 2/3$ and $m_W = 80.6$ GeV in all our simulations. As mentioned in the introduction, we investigated two Higgs masses, $m_H = 35$ and 80 GeV; the former is physically unrealistic but it gives a stronger first-order transition. To study the scaling, we used three different gauge couplings: $\beta_G = 12$, 20 and 32, corresponding to lattice spacings $aT = 4/g^2\beta_G = 0.75$, 0.45 and 0.28125. The lattice volumes varied from $8^3$ up to $32^3$. For each lattice, we did several runs with different values of $\beta_H$ until the transition point was found. This was then converted into a physical $T_c$ with eqs. (6).

One loop perturbation theory predicts that the transition with the above Higgs masses is only very weakly first order. In fact, using the 1-loop *lattice* effective potential with $\beta_G = 20$, we observe that the double-well structure essential for a first-order transition only appears when the linear dimension $N$ of the lattice is at least $N > 29$ when $m_H = 35$ GeV, and $N > 160$ when $m_H = 80$ GeV. In fig. 1 we present the distribution of the link variable $L = \frac{1}{3N^3}\sum_{x,i}V^\dagger(x)U(x)V(x+i)$, $\Phi = RV$, $R \geq 0$, when $\beta_G = 20$ and $m_H = 35$. The two-peak structure characteristic for a first-order transition is obvious. As expected, when $m_H = 80$ GeV the two-peak structure is much less clear (fig. 2), but still visible.

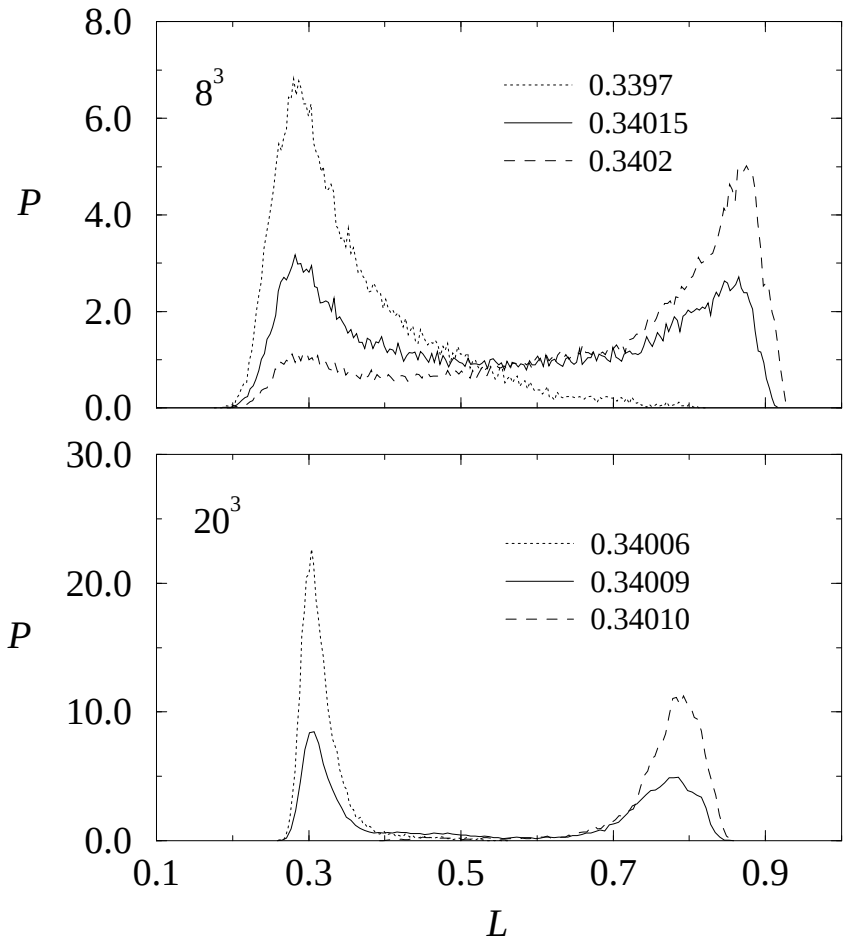

Figure 1. The distribution of the order parameter $L$ when $m_H = 35\,\mathrm{GeV}$ for $8^3$ and $20^3$ lattices for various $\beta_H$-values.

Table 1
The critical temperatures from the simulations and from 1-loop effective potential.

| $m_H$/GeV | $T_c$/GeV | | | |
|---|---|---|---|---|
| | 1-loop | $\beta_G = 12$ | 20 | 32 |
| 35 | 95 | 87(1) | 85(1) | 82(1) |
| 80 | 183 | 158(2) | 157(2) | – |

In table 1 we present the measurements of the critical temperature, together with the 1-loop effective potential results. As we can see, the measured $T_c$ is lower than the perturbative $T_c$, and also that it is fairly independent of the lattice spacing $a$.

To conclude with, we have described a method for reducing the 4d SU(2)-Higgs theory into a 3d effective theory with perturbatively calculable coefficients. This theory can be studied with lattice methods. We have presented results from lattice simulations with $m_H = 35$ and $80\,\mathrm{GeV}$, and we observe that: a) the first-order nature is stronger and b) the transition temperature is lower than predicted by the 1-loop effective potential analysis, making the EW baryogenesis viable for $m_H$ up to 85 GeV. These calculations are discussed in detail in [3], and at 2-loop level in [4].

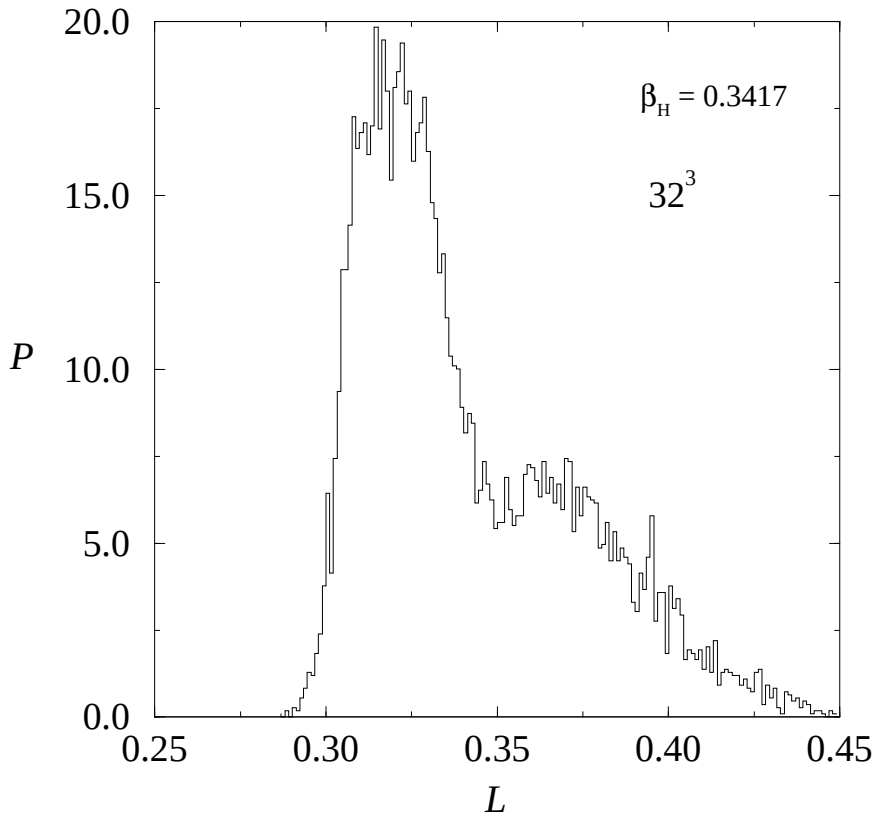

Figure 2. The same as fig. 1 for $m_H = 80\,\mathrm{GeV}$ and $32^3$ lattice.